\documentclass[reprint,
superscriptaddress,
amsmath,amssymb,aps,showkeys,showpacs,
twoside,final,secnumarabic,
nofootinbib,onecolumn]{revtex4-2}

\usepackage[paperwidth=205mm,paperheight=290mm,top=17mm,bottom=25mm,
inner=17mm,outer=17mm,]{geometry}

\usepackage{cmap} 
\usepackage[T1,T2A]{fontenc}
\usepackage[utf8]{inputenc}
\usepackage[russian,english]{babel}
\usepackage{color}
\usepackage{graphicx}
\usepackage{dcolumn}
\usepackage{bm} 
\usepackage[unicode=true,colorlinks=true,linkcolor=magenta, urlcolor=blue, citecolor = blue,breaklinks]{hyperref}
\usepackage{multirow}
\usepackage{url}
\usepackage{breakurl}
\DeclareGraphicsExtensions{.eps}

\newcount\issue
\newcount\Vol
\newcount\numb
\usepackage{fancyhdr} 
\def\numb{x}
\begin{document}

\title{Gamma/hadron separation in the TAIGA experiment\\with neural network methods} 

\def\addressa{Irkutsk State University, Institute of Applied Physics, Irkutsk, 664003, blv. Gagarina, 20, Russia}
\def\addressb{Lomonosov Moscow State University, Skobeltsyn Institute of Nuclear Physics, Moscow, 119991 Leninskie gory,1, bld.2, Russia}

\author{\firstname{E.O.}~\surname{Gres}}
\email[E-mail: ]{greseo@mail.ru}
\affiliation{\addressa}
\affiliation{\addressb}
\author{\firstname{A.P.}~\surname{Kryukov}}
\affiliation{\addressb}
\author{\firstname{P.A.}~\surname{Volchugov}}
\affiliation{\addressb}
\author{\firstname{J.J.}~\surname{Dubenskaya}}
\affiliation{\addressb}
\author{\firstname{D.P.}~\surname{Zhurov}}
\affiliation{\addressa}
\affiliation{\addressb}
\author{\firstname{S.P.}~\surname{Polyakov}}
\affiliation{\addressb}
\author{\firstname{E.B.}~\surname{Postnikov}}
\affiliation{\addressb}
\author{\firstname{A.A.}~\surname{Vlaskina}}
\affiliation{\addressb}

\begin{abstract}
In this work, the ability of rare VHE gamma ray selection with neural network methods is investigated in the case when cosmic radiation flux strongly prevails (ratio up to \(10^4\) over the gamma radiation flux from a point source). This ratio is valid for the Crab Nebula in the TeV energy range, since the Crab is a well-studied source for calibration and test of various methods and installations in gamma astronomy. The part of TAIGA experiment which includes three Imaging Atmospheric Cherenkov Telescopes observes this gamma-source too. Cherenkov telescopes obtain images of Extensive Air Showers. Hillas parameters can be used to analyse images in standard processing method, or images can be processed with convolutional neural networks. In this work we would like to describe the main steps and results obtained in the gamma/hadron separation task from the Crab Nebula with neural network methods. The results obtained are compared with standard processing method applied in the TAIGA collaboration and using Hillas parameter cuts. It’s demonstrated that a signal was received at the level of higher than \(5.5\sigma\) in 21 hours of Crab Nebula observations after processing the experimental data with the neural network method.
\end{abstract}

\par
\keywords{gamma astronomy, IACT, image recognition, convolutional neural networks, Crab Nebula  \\[5pt]}

\maketitle
\thispagestyle{fancy}


\section{Introduction}\label{intro}

Gamma-ray astronomy is a rapidly developing field of observational astrophysics. Gamma particles are unique because of the absence of charge, so they are not deflected by galactic magnetic fields. Due to this fact the presence (or lack) of gamma radiation while studying the sky can tell about the processes occurring in this region of the sky and indicate the presence of a possible gamma-ray source. The separation of the gamma-ray particle flux from the isotropic background cosmic radiation and reconstruction of it's spectrum allows astrophysicists to study the processes happening in supernova explosions or in the ultra-relativistic particles movement in the accretion disk of a black hole. There are different mechanisms for generating gamma photons, which also affect the quantity and energy of gamma radiation emitted. The flux of ultrahigh energy gamma photons (UHE gamma-quanta, energy range 10-100 TeV) becomes so small that in that energy range it is possible to study the sky in gamma rays only from Earth-based installations\cite{bib1}.

To detect and measure the parameters of gamma rays coming from space scientists are developing new methods, creating various installations and conducting different experiments. In gamma-astronomy and astrophysics detection of UHE gamma-ray photons along with cosmic radiation occurs only from the surface of the Earth, using indirect method of extensive air showers\cite{bib2}. As a result of interaction gamma photon with energies more than 1 GeV (high and UHE) and atmosphere, gamma photon decay into an electron-positron pair which generates a large number of electrons of lower energies when passing through the atmosphere and interacting with surrounding atoms. This electromagnetic cascade process is called an extensive air shower (EAS). During the cascade process thousands of relativistic particles are produced, making indirect observations possible. It is worth noting that cosmic radiation also create EASs, however, due to the fact that the cosmic radiation consists of atomic nuclei (mainly protons), a cascading nuclear process will be observed along with electromagnetic processes. And since in both cases the secondary particles of EAS are relativistic, Cherenkov radiation appears in the atmosphere during the shower development, which can also be recorded along with other components of shower. The spatial distribution of Cherenkov radiation strongly depends on the initial (primary) particle. In the process of EAS from cosmic rays many sub-showers arise due to the large number of rapidly decaying \(\pi^0\), \(\pi^+\), \(\pi^-\)-mesons, leading to significantly larger size of the shower from a charged particle than from gamma photon. Classification of events depending on the type of primary particle is possible based on this fact.

There are several methods to detect Cherenkov light from showers and gamma photons specifically, but in this work all attention will be drawn to Imaging Atmospheric Cherenkov Telescopes (IACTs). Cherenkov telescopes in many experiments have a common design: a spherical mirror and a camera made of photomultipliers (PMTs) at the focus of the mirror. The duration of a Cherenkov flash light is on the order of several nanoseconds, but this is sufficient to obtain a "photo" of the EAS as data for further analysis. The installations that include IACTs are the MAGIC\cite{bib3}, H.E.S.S.\cite{bib4}, VERITAS\cite{bib5} experiments and the developing CTA\cite{bib6} and TAIGA-IACT experiments.

The gamma observatory TAIGA, located in the Republic of Buryatia near Lake Baikal, Russia, is a unique facility that register several components of extensive air showers including Cherenkov radiation using IACTs. TAIGA-IACTs  have a spherical mirror with a diameter of about 4 meters with focal length 4.75 m, as well as a camera with a matrix consisting of 560-600 PMTs with field of view \(9.7^\circ\). At this moment three telescopes are in operation. A more detailed description of the TAIGA and TAIGA-IACT experiments can be found in \cite{bib7}.

The main task of TAIGA-IACT is to separate the flux of UHE gamma photons from the hadron background of cosmic radiation and reconstruct the energy spectrum of gamma radiation from the source observed. To do this, Hillas parameter method\cite{bib8, bib9} are used, where the first and second central moments (\emph{Size}, \emph{Length}, \emph{Width}, etc, so called Hillas parameters) are calculated from the EAS "photo", and  after that events are divided according to these parameters. TAIGA-IACTs detect gamma photons in the energy range of several tens and hundreds TeV where a strong imbalance of gamma and cosmic radiation fluxes occurs. The ratio of the gamma flux to hadron flux is approximately \(1:10^4\) for the Crab Nebula which is used to calibrate and verify different installations in gamma astronomy. In the current situation standard processing methods do not allow us to completely separate events caused by gamma photons from hadron events. At the same time bounds and cuts used are not standardized and different combinations of Hillas parameters and conditions on them lead to different results in the standard method. Along with all of the above the Hilas parameter method is the only method of data processing and analysis in TAIGA-IACT, which makes it essential to explore other ways of data analysis that could solve the existing disadvantages of the standard method.

Deep learning methods are considered as an alternative processing method. We would like to notice that the application of deep learning methods in the TAIGA-IACT experiment has already been considered (\cite{bib10, bib11, bib12}), when the signal stands out well with an imbalance of flows \(1:1000\). In this work, this classification method has been developed and supplemented, which makes it possible to achieve good signal separation accuracy even with a greater imbalance of gamma photon flux. This method has also been tested on experimental data from TAIGA-IACT.

\section{\label{sec:level2}Deep learning method for TAIGA-IACT data classification}

Deep learning is a machine learning method that is based on the use of multi-layer representations (Artificial Neural Networks, or ANNs) trained by a computer on large amount of input data. The basis for this method was the Universal Approximation Theorem and the Perceptron Convergence Theorem, proven at the end of the 20th century, which initiated the development of deep learning methods\cite{bib13, bib14}.

\begin{figure}[b]
\includegraphics{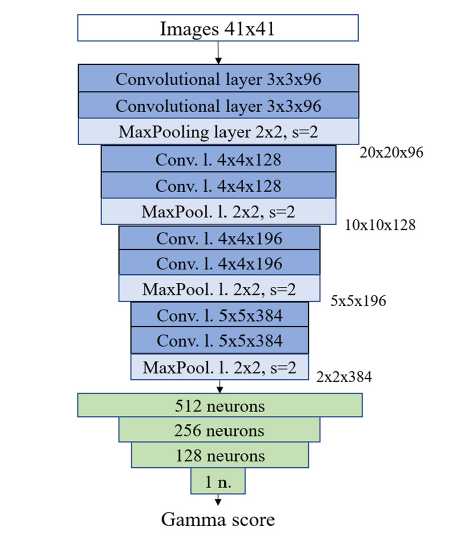}
\caption{\label{fig:fig1} Architecture of convolutional neural network considered in this work.}
\end{figure}

\begin{figure*}
\includegraphics{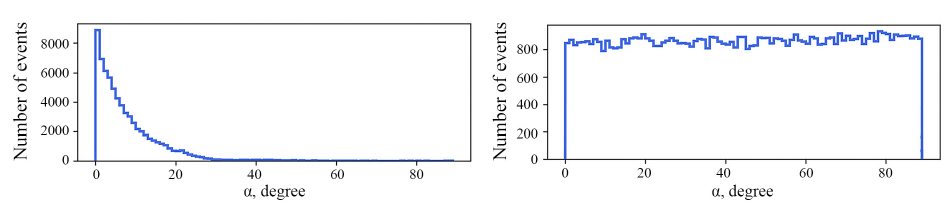}
\caption{\label{fig:fig2} Angle \(\alpha\) distribution for gamma-events (left) and hadrons (right).}
\end{figure*}

One of the first stages in the development of a deep learning method for TAIGA-IACT was the requirement to determine the type and structure of a neural network. Since the input data are images a convolutional neural network (CNN) (Fig.~\ref{fig:fig1}) was considered, consisting of a block of convolutional layers that analyze image and extract structural or statistical features in images for each class and a block with fully connected layers classifying IACT images. By the class in the input data we mean the type of the primary particle that produced air shower: \emph{0} is a hadron, \emph{1} is a gamma photon.

Python was used as a tool for creating CNN using Tensorflow and Keras libraries. \emph{ReLU} (Rectified Linear Unit) activation function was applied between hidden layers in CNN, however, the sigmoid activation function is used at the output. As loss function binary cross-entropy was used in CNN training. Taking into account all the above this determines the classification problem that CNN should solve. The output data are values in the range [0;1], which are interpreted as measure of the probability that the image under consideration by CNN is generated by gamma photon (the so-called gamma score of the image).

\subsection{\label{sec:level21}Training data set}

An equally important aspect in the neural network method is the data on which the training takes place. It is due to the fact that the neural network learns to approximate on the training samples, therefore the training data must be large and representative, i.e. contain all the expected phenomena in the experiment. Thus, the training dataset included Monte Carlo (MC) simulated gamma events\cite{bib15}, the spectrum of which had a spectral index -2.6 in the energy range from 1 to 200 TeV (set was compiled to approximately correspond to the spectrum of the Crab Nebula in this energy range). Gamma images were generated using CORSIKA software\cite{bib16} and specialized software for simulation of the TAIGA-IACT response\cite{bib17}.

The simulation was applied only to create a data of the class \emph{1}. Experimental data (hadrons) were taken for the class \emph{0}. The decision to use experimental data instead of MC protons is due to the fact that the correct spectrum and nuclear composition of cosmic radiation are already taken into account in the experiment naturally. In the experimental data there are also possible noise effects related with the operation of the telescope that can be hardly predict in simulation.

Before training all images were preprocessed as follows: cleaning, image modification according Wobble pointing mode, image transformation to a square shape and logarithmic scaling of image pixels. The cleaning step involves removing islands of noise pixels created by background illumination (the low cleaning threshold was 7 phe, the high threshold was 14 phe). Image modification according Wobble pointing mode\cite{bib18} considers because in TAIGA simulations all events are generated with the source of gamma photons in the center of the camera, which is not true in the experiment. Due to the Wobble the gamma-ray source is located at a certain distance from the center of the camera, which, as shown by the experiment in \cite{bib12}, can be a critical in image recognition and classification by neural networks in TAIGA-IACT. Also Wobble helps to estimate the number of false-positive events. Image transformation to a square shape is necessary because the original image structure is hexagonal, and CNN can not process such images. To bypass this restriction of CNNs, the image transformation algorithm was performed using the axial method \cite{bib19}. The logarithmic scaling of pixel amplitudes which converts amplitudes into the range of values [0;1] allows us to improve the quality of CNN training.

In previous works \cite{bib10, bib11, bib12} it was established that despite the studied neural network structures the achievable limit of hadron background suppression \(B\), defined as the ratio of the numbers of hadrons before and after classification, is no more than 1000. In order to increase \(B\) to the desired value of 10000, it was decided to sort the data by some substantial feature, which distribution shape would be very different from the type of the primary particle. The Hillas parameter, the angle \(\alpha\), was considered as such substantial feature. This angle determines the orientation of major axis of the ellipse, which describes the spot of EAS on the camera, relatively the gamma source. Figure~\ref{fig:fig2} shows the distributions of the angle \(\alpha\) depending on the type of primary particle. This effect is related to the physics of the process, since cosmic radiation flux is isotropic (this statement is true for particle energies up to \(10^{18}\) eV). Gamma photons being neutral particles come almost directly to Earth from a gamma-ray source, creating a pronounced peak in the orientation of the ellipses on the telescope camera. Together with the angle \(\alpha\), total pixel brightness (\(Size\)) was also considered to reject events with \(Size\) less than 120 phe. The rejection of such events is necessary to avoid strongly distorted with cleaning or too dim images.

Taking into account all the above, MC analysis showed that after \(Size > 120\) phe and \(\alpha < 20^\circ\) approximately \(90\%\) gamma photons and \(20\%\) hadrons remain (next will be referring as soft selection conditions). After \(Size > 120\) phe and \(\alpha < 6^\circ\) there remain \(50\%\) gamma photons and less than \(10\%\) hadrons (strict selection conditions). Consequently, the hadron background is suppressed already 10 times with this approach with minimal loss of gamma events, so CNN was trained on a training dataset with soft selection conditions. Training set consisted of 38400 MC gamma events and 40000 experimental hadron events. To enlarge the number of training samples artificial expansion was applied using image rotations every \(60^\circ\)  because of the symmetry of the hexagonal structure of the camera matrix.  Therefore the number of training samples was equal to 470400. The training lasted for 25 epochs to avoid CNN overfitting.

\subsection{\label{sec:level22}Validation}

\begin{figure}[b]
\includegraphics{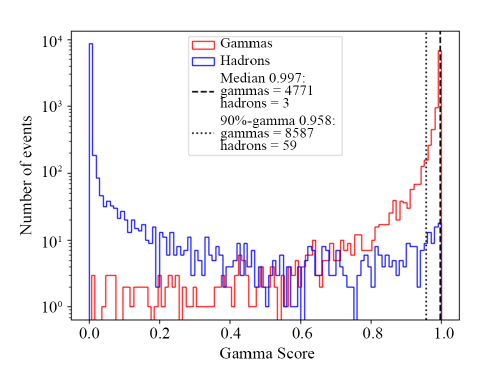}
\caption{\label{fig:fig3} Distribution of CNN classifier output for validation data, consisting of 9500 MC gamma events and 9600 experimental hadrons.}
\end{figure}

Validation helps to estimate the quality of the classifier's training. The validation set with soft selection conditions consisted of 9500 MC gamma events with the same energy characteristics as the gamma events in the training set and 9600 experimental hadrons. All events have been preprocessed in the same way as the training samples. The distribution of gamma score for gamma and hadron events of validation set after training are demonstrated in Figure~\ref{fig:fig3}. Since as an output gamma score lies in range [0;1], it's necessary to determine only two classes - gamma photon (1) or handron (0). To make this, class threshold needs to be defined. In gamma astronomy separation threshold usually are taken at \(50\%\) gamma photon loss from the primary set (in this case validation set). Consequently, that leads to the value of class threshold of 0.9965, when 4771 gamma photons and 3 hadrons remain. In this case the accuracy of the classifier is \(99.93\%\) and the hadron suppression \(B\) is approximately 3000. Taking into account the soft selection conditions, the classification quality estimation shows that the coefficient \(B\) reaches the desired value of 10000 with a loss of approximately \(50\%\) gamma events.

\section{\label{sec:level3}Test verification on experimental data and comparison with standard processing method}

\begin{figure*}
\includegraphics{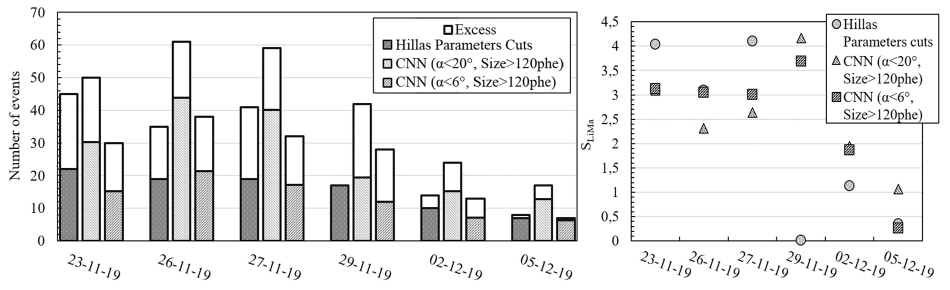}
\caption{\label{fig:fig4} Results obtained in the classification task to separate signal from the Crab Nebula using standard and deep learning methods. On the left picture diagram with average hadron background (calculated in 10 OFF points; filled columns) and excess in ON point at each observation night (empty columns) are shown. Right figure demonstrates the calculated signal significance at each night using LiMa calculation\cite{bib20}. }
\end{figure*}

Since the desired quality of hadron suppression was achieved on the validation set, the next step was to apply the neural network method to the test set of experimental data. For this purpose, 6 nights were selected in November-December 2019, when the Crab Nebula was observed. The Crab nebula is a well-studied source in many spectral ranges, including the gamma ray with energies of the order of 10 TeV, therefore this source is used in calibrations and testing of new observation methods in gamma astronomy. Before classification events were taken in clear nights, when the weather index, based on the estimation of the upper limit of magnitude with CCD camera, was greater than 9, and the count rate of cleared events was greater than 8 Hz.

The LiMa significance (\cite{bib20}, Eq. 17) calculation was used to estimate the level of extracted signal. To assess the number of the hadron background on the camera, according to the Wobble pointing mode, 10 OFF-points were taken with one ON point, where the Crab Nebula was located when pointing.

Alongside with the deep learning method experimental data were also processed using the Hillas parameter method. It was mentioned that the standard method is based on the calculation of the first and second moments from images, which, from a purely geometric point of view, parameterize the spot on the PMT matrix with an ellipse. Parameters such as  \emph{Length} and \emph{Width} are the major and minor axes of the ellipse respectively, \emph{Size} is the total brightness of the image, \emph{Distance} and angle \(\alpha\) or \(\theta^2\) define the distance and orientation between the center of gravity of the ellipse and the center of the camera matrix (after taking into account Wobble pointing method).

For the TAIGA-IACT experiment, the following restrictions (cuts) on the Hillas parameters were derived, which successfully allowed us to isolate the signal from the Crab Nebula at the level of 12\(\sigma\) using LiMa estimation of signal significance for 150 observation hours\cite{bib21}:
\begin{eqnarray}
&Size> 120 phe; 0.36^\circ< Distance < 1.44^\circ;&\nonumber\\
&\label{eq:one}0.024<Width< (0.068^\circ*lg(Size)-0.047^\circ);&\\
&Length < (0.145^\circ*lg(Size)); \theta^2<0.05^\circ. &\nonumber
\end{eqnarray}
With further explanations in this work Hillas parameters cuts refer to the set of bounds and dependencies specified in Eq.~(\ref{eq:one}).

\begin{figure*}
\includegraphics{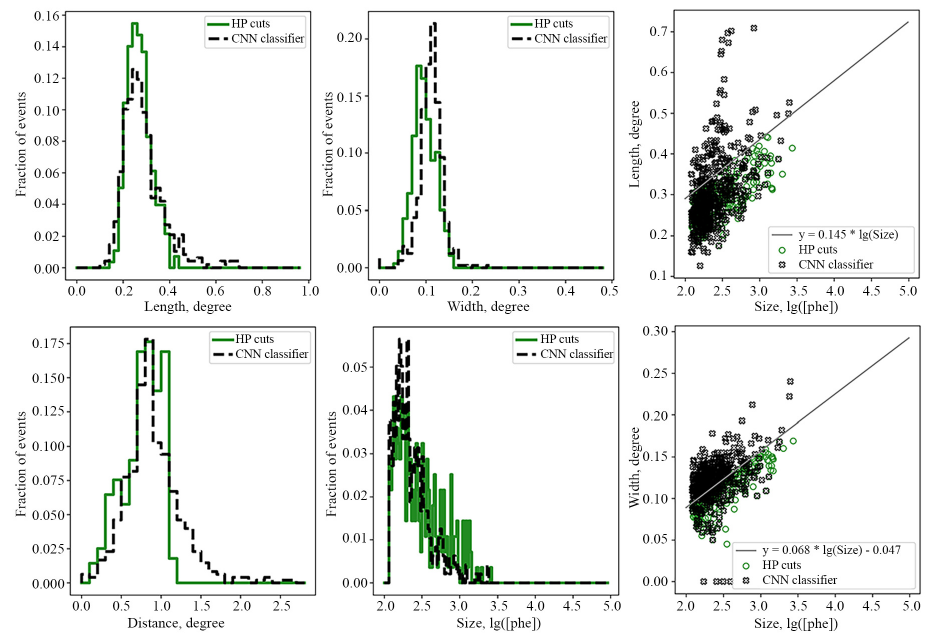}
\caption{\label{fig:fig5} Distributions and interdependence of Hillas parameters for events selected at ON point by standard and CNN with strict selection conditions during observation of the Crab Nebula for 21 hours.}
\end{figure*}

\begin{table*}[htbp]
\renewcommand{\arraystretch}{1.25}
\renewcommand{\tabcolsep}{3pt}
\begin{center}\caption{Classification results and assessment of signal significance with standard and deep learning methods on Crab Nebula data obtained at November-December 2019 year.}
\begin{tabular}{|c|c|c|c|c|c|}\hline
Processing method                                                       & Observation time, hour  & \(<N_{OFF}>\) & \(N_{ON}\) & \(Excess\) & \(S_{LiMa}\) \\ \hline
Hillas Parameters cuts                                                  & \multirow{3}{*}{21}     &   94  &  160   &     66   &   5.84  \\ \cline{1-1}\cline{3-6}
CNN with \(Size > 120\) phe and  \(\alpha < 20^\circ\)  &   {}                             &   161.8  &  253   &  91.2   &   6.26  \\ \cline{1-1}\cline{3-6}
CNN with \(Size > 120\) phe and  \(\alpha < 6^\circ\)    &   {}                             &   79.3   &  148   &  68.7   &   6.48  \\ \cline{1-1}\cline{3-6} \hline
\end{tabular}\label{tab:table1}
\end{center}
\end{table*}

Figure~\ref{fig:fig4} demonstrated the average hadron background \(<N_{OFF}>\) calculated from 10 OFF points, excess over background in ON point where source is located, and LiMa criterion calculated for each data analysis method on each night. We would like to note that in the first 4 nights under consideration the observations were quite long (3-4 hours), unlike on 2 and 5 December (1-2 hours). This affects the total number of events at both the OFFs and ON points when counting events. Also it's easy to see from the diagram that strict selection conditions in CNN method allows us to achieve the same level of the average background as in the Hillas parameter method with low losses in the excess. That is why two selection conditions, soft and strict, were considered for separating events before giving them to CNN. However, the complete separation was not reachable with CNN method despite two selection conditions. It can also be seen that the excess and \(S_{LiMa}\) are significantly different for one night (29 November 2019). Since the data was already selected by weather conditions, it is not entirely clear why there is such a strong difference in classification on one night while on other days everything is approximately the same.

Table~\ref{tab:table1} shows the assessment for all nights in total. It is demonstrated that the quality of CNN classification, regardless of the selection conditions, is on the same level than the standard method. In the meantime, 46 events were selected simultaneously by the standard method and CNN (soft selection conditions) in ON point. However if we exclude 29 November from consideration, we get that the signal from the Crab Nebula is allocated at the level of 6.3\(\sigma\), 5.1 and 5.5\(\sigma\) by the Hillas parameters and CNN methods with soft and strict selection conditions, respectively. In this case, the standard method gives a higher significance level than the CNN method, so fine-tuning and adjustment of the model used in this work is still required to improve the efficiency of CNN in the experiment.

Since neural network is essentially a black box, it is difficult to track the dependence of millions of parameters with complex nonlinear connections between layers. Therefore, it was also necessary to compare the selected events using those geometric parameters (the same Hillas parameters), the distributions of which a person is able to investigate and understand. The distributions and dependencies of Hills parameters for events selected by both methods (standard method and CNN classifier with strict selection conditions) of analysis at the ON point of observation are demonstrated in Figure~\ref{fig:fig5}. For CNN with soft selection conditions the shape of the distributions are similar to the ones showed. It can be seen from the graphs that the distributions of parameters such as \emph{Size}, \emph{Distance} and \emph{Length} are in good agreement with each other and the same boundaries under study (see Eq.~(\ref{eq:one})). \emph{Width} distributions are an exceptions. Neural network method, along with narrow ellipsoid events, selected events with large values of minor axes \emph{Width}, making almost half of the events be above the linear dependence of \emph{Width(Size)}. Since CNN is an approximator with complex nonlinear connections within itself, it may happen that distributions under consideration are not precisely coincided.

\section{Conclusion}\label{concl}

The gamma observatory TAIGA  is a unique installation with many detectors and tools for searching for answers in many areas of theoretical physics. In gamma ray astronomy, one of the main observation instruments  which uses various methods to separate gamma photons from cosmic ray radiation and to reconstruct their primary characteristics is the Imaging Atmospheric Cherenkov Telescope.

In the course of this work the method of convolutional neural networks to solve problems in processing and analyzing data from TAIGA-IACTs was considered. Despite the fact that the complete separation of gamma photons and hadrons in the experiment was not achieved, it was shown that deep learning methods give the result comparable to the standard method of data processing and analysis. When studying the quality of classification and comparison with standard processing method, it can be concluded that CNN can be an effective and independent selection approach at this stage of the work. In particular, the signal from Crab Nebula were obtained on 6.5\(\sigma\) level of significance for 21 hours of observation time. 68 gamma events were obtained during this period of time.

In the future, it is planned to further develop and improve neural network methods for processing and analyzing experimental TAIGA-IACT data. The next stages of the work are to consider a larger amount of data to gather more statistics and, along with this, to study other gamma sources observed by TAIGA-IACT using this method. Since signal extraction is the first step in analyzing experimental data in gamma ray astronomy, in perspectives it is also planned to develop deep learning methods to reconstruct the energy spectrum from experimental data.

\begin{acknowledgments}
The authors would like to thank the TAIGA collaboration for support and
data provision. The work was carried out using equipment provided by the
MSU Development Program.
\end{acknowledgments}

\nocite{*}

\section*{FUNDING}
This study was supported by the Russian Science
Foundation, grant no. 24-11-00136.


\end{document}